%% file: nspe.tex
\documentclass[fleqn,usenatbib]{mnras}
\usepackage{newtxtext,newtxmath}

\usepackage[T1]{fontenc}
\newcommand\thefont{\expandafter\string\the\font}
\DeclareRobustCommand{\VAN}[3]{#2}
\let\VANthebibliography\thebibliography
\def\thebibliography{\DeclareRobustCommand{\VAN}[3]{##3}\VANthebibliography}

\usepackage{graphicx}	
\usepackage{amsmath}	

\title[Nested sampling parameter estimation]{Costless correction of chain based nested sampling parameter estimation in gravitational wave data and beyond}

\author[]{
Metha Prathaban$^{1,}$$^{2,}$$^{3}$\thanks{E-mail: myp23@cam.ac.uk}
and Will Handley$^{1,}$$^{2,}$$^{4}$\thanks{E-mail: wh260@cam.ac.uk}
\\
$^{1}$Kavli Institute for Cosmology, Madingley Road, Cambridge CB3 0HA, UK\\
$^{2}$Astrophysics Group, Cavendish Laboratory, J.J. Thomson Avenue, Cambridge CB3 0HE, UK\\
$^{3}$Pembroke College, Trumpington Street, Cambridge CB2 1RF, UK \\
$^{4}$Gonville \& Caius College, Trinity Street, Cambridge CB2 1TA, UK
}

\date{Accepted XXX. Received YYY; in original form ZZZ}

\pubyear{2024}

\begin{document}
\label{firstpage}
\pagerange{\pageref{firstpage}--\pageref{lastpage}}
\maketitle

\begin{abstract}

Nested sampling parameter estimation differs from evidence estimation, in that it incurs an additional source of uncertainty. This uncertainty affects estimates of parameter means and credible intervals in gravitational wave analyses and beyond, and yet, it is typically not accounted for in standard uncertainty estimation methods. In this paper, we present two novel methods to quantify this uncertainty more accurately for any chain based nested sampler, using the additional likelihood calls made at runtime in producing independent samples. Using injected signals of black hole binary coalescences as an example, we first show concretely that the usual uncertainty estimation method is insufficient to capture the true error bar on parameter estimates. We then demonstrate how the extra points in the chains of chain based samplers may be carefully utilised to estimate this uncertainty correctly, and provide a way to check the accuracy of the resulting error bars. Finally, we discuss how this uncertainty affects $p$-$p$ plots and coverage assessments.
\end{abstract}




\section{Introduction}

Nested Sampling (NS)~\citep{skilling} is a Bayesian inference tool widely used in the field of gravitational wave astronomy, both for parameter estimation and model comparison~\citep{nature_ns_review, buchner_ns_review, Veitch_2010, Thrane_2019, nessai, multinest, lal, bilby}. It enables us to perform posterior and evidence estimation on observed gravitational wave events, in turn enabling breakthrough science, such as an independent measurement of the Hubble constant~\citep{hubble_theory, hubble}, reconstructed sky maps of binary neutron star mergers for electromagnetic follow-up~\citep{skymap}, and population inference to understand formation mechanisms of binaries~\citep{population1,population2, population_review}.

Nested sampling distinguishes itself from Markov Chain Monte Carlo (MCMC) algorithms~\citep{MacKay2003, emcee, bilby-mcmc} in its ability to easily calculate evidences, and the uncertainties associated with this evidence estimation are well understood~\citep{skilling}. Whilst it is also popularly used for parameter inference in gravitational wave data analysis and beyond, nested sampling parameter estimation contains an additional source of uncertainty which affects its performance~\citep{Chopin_Robert}. Quantifying this uncertainty accurately is key, as it affects our estimates of parameter means and credible intervals on parameters. This uncertainty is essentially a result of the stochastic nature of the nested sampling algorithm, and so we are able to accurately quantify it by performing many nested sampling runs on the same data. However, this is computationally expensive and, for many gravitational wave problems, simply not viable. There is, as yet, little literature exploring accurately estimating this uncertainty from a single run. In response to this,~\cite{Higson_2018} proposed an innovative method of deconstructing nested sampling runs into single live point runs and using bootstrapping to determine a new error bar on parameter inferences. 

Here, we present two novel approaches which utilise the extra likelihood evaluations already performed at runtime for any chain based nested sampler. Deemed to be too correlated to use in evidence estimation, the potential performance gains to parameter inference from this wealth of additional likelihood calls has been largely unexplored. As an example of this, we show that, by carefully harnessing these additional evaluations, we can correctly quantify parameter estimation uncertainties and verify that these match the uncertainties obtained from repeated nested sampling runs. 

In the following section, we lay out necessary background on the nested sampling algorithm and identify the dominant sources of uncertainty for both evidence and parameter estimation. In Section~\ref{gwsim}, we first verify empirically, in the context of a simulated black hole binary (BBH) signal, that the standard methods of uncertainty estimation in nested sampling do not accurately capture the true error bars on parameter estimates. We then introduce the two new methods. Section~\ref{results} demonstrates the application of these methods to infer the correct uncertainties on parameter means and credible intervals, and show that they produce comparable results to the method presented in~\cite{Higson_2018}. Finally, our conclusions are presented in Section \ref{conc}.  

\section{Background}

\subsection{Anatomy of a nested sampling run}

A typical nested sampling run begins by populating the parameter space with a number of `live points', drawn from the prior, and evaluating the corresponding likelihood values of these points. At each iteration of the algorithm, the live point with the lowest likelihood is deleted and a new point is drawn from the prior, with the constraint that it must have a likelihood higher than that of the deleted point. As this process continues, the collection of live points compresses exponentially in the parameter space towards the peak of the likelihood function~\citep{aeons} until some termination condition is satisfied. 

\begin{figure}
    \centering
    \def\svgwidth{\columnwidth}
    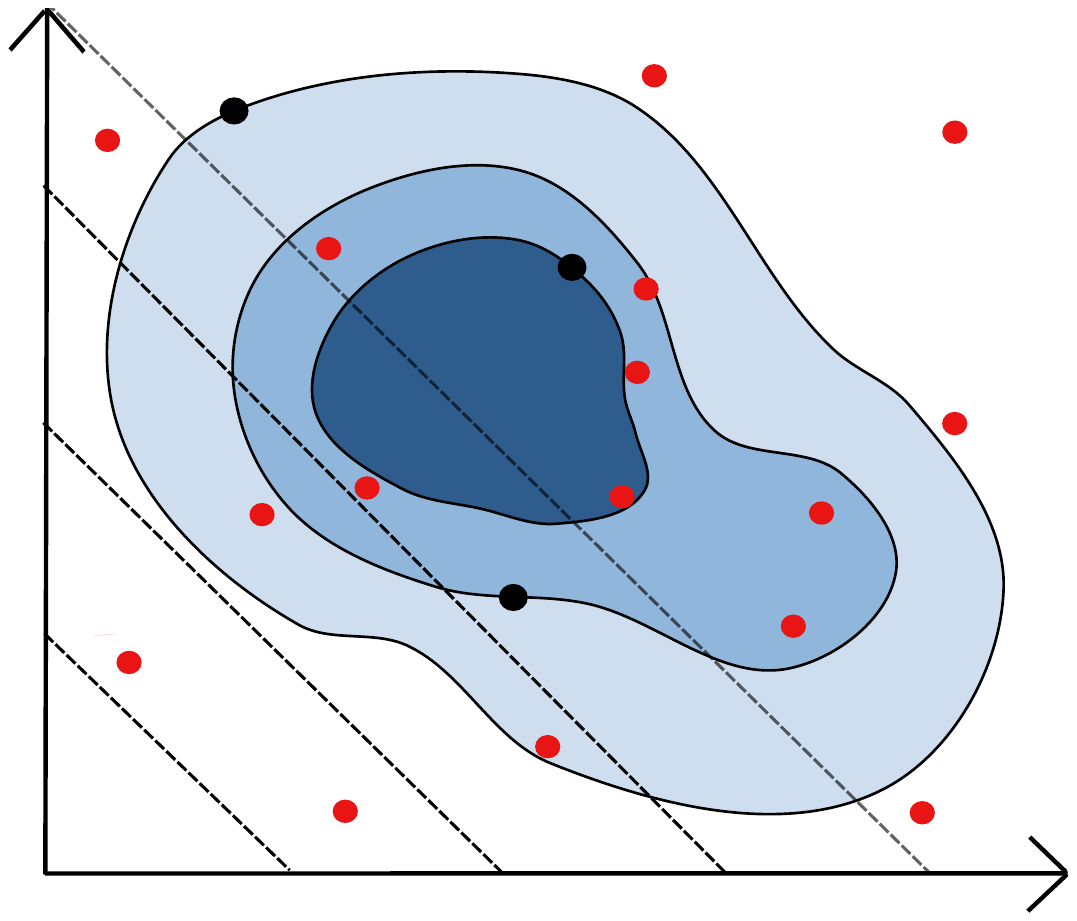
    \vspace{-125pt}
    \caption{Schematic of a typical nested sampling run with a chain based sampler. At the end of the run, we are left with a set of dead points (black), which define a series of nested iso-likelihood contours. To generate a new live point from a given point, chain based samplers use a Markov-Chain based procedure to continually generate points within its likelihood contour, until the new point is deemed uncorrelated enough with the original point from which it was seeded. Thus, at the end of the run we are also left a set of `phantom points' (red), the exact number of which depends on the chain lengths. An example chain is shown in black between dead points $2$ and $3$. In parameter estimation, we typically only use the dead points and, therefore, must use the parameter values of each dead point as a proxy for the average parameter value along the entire contour. For an example two-parameter case, where the parameter being estimated is the sum $f(\theta)=\theta_1 + \theta_2$, the contours of constant of constant $f(\theta)$ are shown (dashed). In this case, the parameter value of the dead points is not necessarily representative of the average parameter value over the contours, and this will be the dominant source of uncertainty in our $f(\theta)$ estimate. However, the phantom points can provide a better understanding of the variation of this parameter along the contours, enabling a more accurate quantification of this uncertainty.}
    \label{fig:schematic}
\end{figure}

At the end of the run, we are left with the series of deleted points, termed `dead points', each associated with a set of parameter values, $\theta_i$, and a likelihood, $\mathcal{L}_i$ (see Figure~\ref{fig:schematic}), and each defining an iso-likelihood contour in the parameter space. Every dead point is also assigned a value $X_i$, defined as the fractional prior volume contained within the iso-likelihood contour defined by the point:
\begin{equation}
   X(\mathcal{L}_i) \equiv \int_{\mathcal{L}(\theta) > \mathcal{L}_i} \pi(\theta) d\theta. 
\end{equation}
By construction, $X$ runs from 1 to 0. The exact prior volumes are unknown, but are modelled statistically using a set of shrinkage ratios, $\textbf{t}$, where $X_i = t_i X_{i-1}$. $t_i$ is defined as the largest of $N$ random numbers drawn from a uniform distribution between zero and unity~\citep{skilling} and, thus,
\begin{equation}
    P(t_i) = Nt_i^{N-1}. 
\end{equation}

\subsection{Evidence estimation}

In the subsequent sections of the background, we will describe the different sources of uncertainty present in evidence estimation and parameter estimation from a nested sampling run. Much of this argument follows~\cite{Higson_2018}. 

Bayes' theorem tells us that the evidence may be computed by an integral over our parameters as
\begin{equation}
    \mathcal{Z} = \int \mathcal{L}(\theta) \pi(\theta)d\theta.  
\end{equation}
Nested sampling enables the conversion of this many-dimensional integral into a one-dimensional integral, by changing the integration variable to the fractional prior volume within an iso-likelihood contour, $X$:
\begin{equation}
    \mathcal{Z} = \int_{0}^{1} \mathcal{L}(X)dX.
\end{equation}
In practice, this integral is approximated by a sum over the dead points:
\begin{equation}
    \mathcal{Z}  = \int_{0}^{1} \mathcal{L}(X)dX \approx \sum_{i\in \textrm{dead points}} \mathcal{L}_i \Delta X_i.
\end{equation}
By construction, using the likelihood value of a single dead point, $\mathcal{L}_i$, as a proxy for $\mathcal{L}(X)$ is an exact substitution. However, the exact values of the fractional prior volume `shells', $\Delta X_i$, are unknown, since the exact set of shrinkage ratios, $\textbf{t}_i$, are unknown, and this is the dominant source of uncertainty in evidence estimation. Typically, the resulting error bar on the evidence is quantified by simulating sets of $\Delta X_i$ to use in the evidence calculation and quoting the standard deviation from this set of estimates.

\subsection{Parameter estimation}

To calculate the expected value of some function, $f(\theta)$, of the parameters, we must integrate the function over the posterior~\citep{Chopin_Robert}:
\begin{equation}\label{eq:peint}
    E[f(\theta)] = \int f(\theta) \frac{\mathcal{L}(\theta) \pi(\theta)}{\mathcal{Z}} d\theta = \frac{1}{\mathcal{Z}} \int \tilde{f}(X) \mathcal{L}(X)dX.
\end{equation}
$\tilde{f}(X)$ represents the prior expectation of $f(\theta)$ given that $\mathcal{L}(\theta) = \mathcal{L}(X)$:
\begin{equation}
    \tilde{f}(X) = E_{\pi|\mathcal{L}(\theta)=\mathcal{L}(X)}[f(\theta)].
\end{equation}
In order words, it can be seen as the average value of $f(\theta)$ over the given iso-likelihood contour. Discretising~\ref{eq:peint}, as before, gives:
\begin{equation}\label{eq:pe_sum}
    \frac{1}{\mathcal{Z}} \int \tilde{f}(X) \mathcal{L}(X)dX \approx \frac{1}{\mathcal{Z}} \sum_i \tilde{f}(X_i) \mathcal{L}_i\Delta X_i.
\end{equation}
Again, the exact prior volumes are unknown, but there is an additional source of uncertainty here which was not present before: we are required to use a single $f(\theta)$ value on each contour, $f(\theta_i)$, as a proxy for $\tilde{f}(X_i)$. Unlike with the likelihood, it is no longer true that this is an exact substitution, and in parameter estimation this becomes the dominant source of uncertainty. The stochasticity of nested sampling means that over a single run we will not necessarily be able to get representative values of $f(\theta)$ along every contour, and thus won't be able to capture our true error bar over multiple runs, even when simulating sets of prior volumes. This is demonstrated in Figure~\ref{fig:schematic}, where we show the example of trying to estimate the sum, $f(\theta) = \theta_1 + \theta_2$, of parameters in a simple two-parameter case; here, the values of $f(\theta)$ at each of the dead points on the contours do not reflect the average value over the contours. The key to capturing this uncertainty is better understanding the variation of $f(\theta)$ along the contours.

\section{Methods}\label{gwsim}

\begin{figure}
    \centering
    \includegraphics{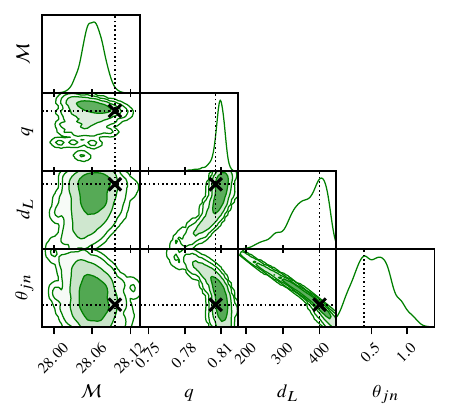}
    \caption{Posteriors recovered from the injected signal using \textsc{PolyChord}, with the injected parameter values indicated. Since parameter estimation studies require large numbers of runs, we only sample these $4$ parameters, with the other parameters simply set to their injected values.}
    \label{fig:posteriors}
\end{figure}

In order to study the nested sampling parameter estimation uncertainty, we consider the example of a simulated black hole binary (BBH) signal with parameters similar to those of GW150914 and a signal-to-noise ratio (SNR) of 51. The exact injected parameter values are shown in Figure~\ref{fig:posteriors}. Since studying the error bars carefully requires hundreds of nested sampling runs to be performed, we work within a simplified framework in which only four of the parameters are sampled: the chirp mass ($\mathcal{M}$), mass ratio ($q$), luminosity distance ($d_L$), and zenith angle between the total angular momentum and line of sight ($\theta_{jn}$). For the results in the following subsection, and for those in section~\ref{results} unless otherwise specified, we perform nested sampling runs with $500$ live points per run and the default \textsc{PolyChord} chain length of $5*\textrm{ndims}=20$. All other sampler settings were set to the \textsc{PolyChord} defaults. The runs are performed using \texttt{bilby}~\citep{bilby}, with a modified version of the in-built \textsc{PolyChord} sampler option. All waveforms are both injected and sampled with the \texttt{IMRPhenomPv2} waveform model~\citep{IMRPhenomP} and we use two interferometers, LIGO-Hanford (H1) and LIGO-Livingston (L1). 

\subsection{Evidence uncertainty estimation method is insufficient for parameter estimation}\label{demo}

\begin{figure*}
    \centering
    \includegraphics{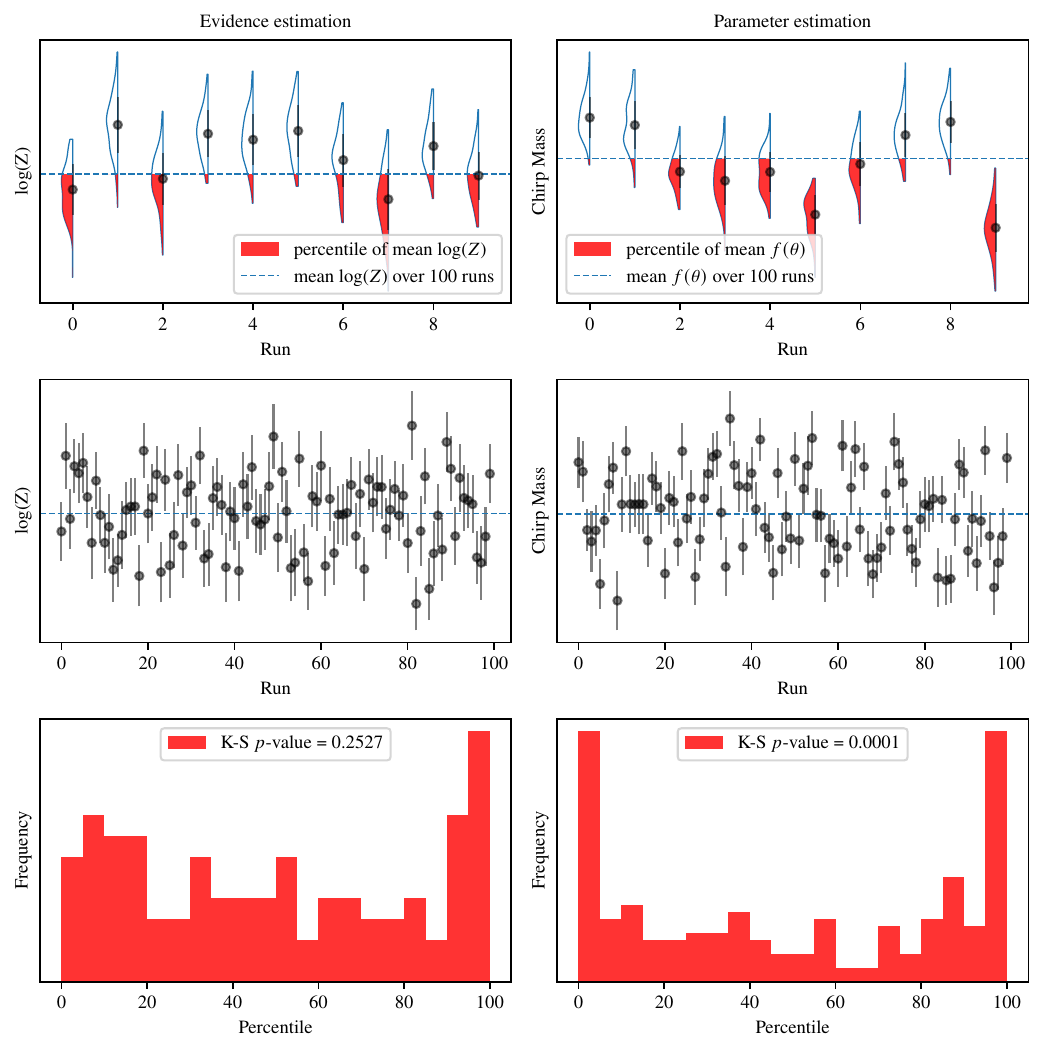}
    \caption{Typically, a distribution of evidence estimates is computed from a single run using the `simulated weights' method (top left). The mean and standard deviation of this distribution are then quoted as the evidence value and its corresponding uncertainty (black). If the error bars from single runs accurately quantify the uncertainty on the evidence due to unknown prior volumes, the overall mean evidence, computed over many runs, should lie in the $1^{\textrm{st}}$ percentile of estimates $1\%$ of the time, in the $50^{\textrm{th}}$ percentile $50\%$ of the time, and so on. Hence, plotting a histogram of the percentiles from each run in which the overall mean (blue dashed line) lies should give a uniform distribution from 0 to 100 (bottom left). We see here that this is indeed the case for the evidences, with the Kolmogorov-Smirnov $p$-value being 0.25, demonstrating that resampling the prior volumes is sufficient for estimating the error bar on the evidence for this example. 
    For the chirp mass, even by eye the estimates per run are not wide enough to capture the variation in estimates between runs (top and middle right). This indicates that the variation in chirp mass along a given contour is not being properly accounted for. The percentiles plot (bottom right) confirms that the uncertainty estimate on a single nested sampling run is too optimistic. More often than would be the case for correctly distributed errors, the overall mean estimate for the chirp mass over 100 runs (blue dashed line) lies deep in the tails of the estimated chirp mass from a single run. The simulated weights method is not sufficient for calculating the parameter estimation uncertainty.}
    \label{fig:combined_violins}
\end{figure*}

The stochasticity of the fractional prior volumes in nested sampling means that the estimated mean evidence will vary from run to run. In order to accurately quantify the uncertainty on a single run, therefore, the error bar must reflect this run-to-run variation. For evidence estimation, this is typically achieved by sampling many sets of the shrinkage ratios, $\textbf{t}$, from a known distribution and using these to calculate the evidences, quoting the error bar from the spread of these estimates.~\cite{Higson_2018} term this the `simulated weights method' and we shall refer to it henceforth as such too. 

In the context of our simulated BBH signal, we can carefully verify that this method does indeed quantify the evidence uncertainty correctly. To do this, we first perform $100$ nested sampling runs on our simulated signal. For each run, we apply the simulated weights method using \texttt{anesthetic}~\citep{anesthetic} to calculate a set of evidence estimates. The resulting distributions of evidences are plotted in Figure~\ref{fig:combined_violins}. The dashed line represents the overall mean evidence calculated from all $100$ runs. 

The errors should be normally distributed, meaning that about $68\%$ of the time our `true evidence' (estimated from the mean of $100$ runs) should lie within the error bars, and $95\%$ of the time it should lie within two times the extent of the error bars. This condition can also be viewed in terms of the distribution of the percentiles of the `true evidence'. If the errors are indeed correctly distributed, a histogram of these percentiles will yield a uniform distribution from $0$ to $100$. For our simulated BBH example, this histogram is plotted in Figure~\ref{fig:combined_violins}, along with the Kolmogorov-Smirnov (K-S) test $p$-value for a uniform distribution. We can conclude from this that the simulated weights method leads to correct evidence uncertainty estimates from a single run.


~\cite{skilling} suggests using the same method to estimate parameter estimation uncertainties. As an example, we shall attempt to estimate the chirp mass of our signal. As before, for each of the $100$ runs, we can simulate sets of shrinkage ratios to obtains sets of $\Delta X$ and substitute these into equation~\ref{eq:pe_sum} to calculate a set of chirp mass estimates. These are plotted in Figure~\ref{fig:combined_violins}, along with the overall mean chirp mass estimated from all $100$ runs.

This time, even by eye it seems that the resulting distributions of chirp masses per run are not quite wide enough to capture the run-to-run variation. Performing a similar analysis of the percentiles as for the evidences yields the histogram in Figure~\ref{fig:combined_violins}. It is unmistakable from both the plot and the K-S test $p$-value that the simulated weights method alone does not lead to correct uncertainty estimates on the chirp mass. Far more often than ought to be the case, the mean chirp mass over many runs lies deep in the tails of the estimates from a single run. This is an empirical verification, in the context of gravitational waves, of results already established theoretically by~\cite{Chopin_Robert} and experimentally by~\cite{Higson_2018}. 

\subsection{Phantom points}\label{phantom}

In nested sampling, there are many ways to generate a new live point subject to the hard likelihood constraint $\mathcal{L} > \mathcal{L}_i$, and this is a key point of difference between existing nested sampling implementations. However, many implementations use a Markov-Chain based procedure, where new points are continually generated within the likelihood contour until we are satisfied that the new point is independent from the deleted point from which it was seeded. This point is then assigned as the new live point, and the points generated in the chain between the deleted and new live point are typically discarded. These points are shown in red in Figure~\ref{fig:schematic}, with an example chain between a deleted and new live point illustrated by the black lines. 

Though deemed too correlated to the deleted point to use as our new live point, these discarded points still have the potential to provide useful information about the parameter space, though this potential has been largely unexplored. For the remainder of the paper we shall refer to these `intra-chain' points as `phantom points', following the terminology of \textsc{PolyChord}~\citep{polychord, polychord2}, the nested sampling implementation we use to obtain the results for this paper; tailored for high-dimensional parameter spaces and parallelised using \textsc{openMPI}, it is particularly suitable for gravitational wave parameter inference studies. However, we emphasise that the existence of these phantom points is not unique to \textsc{PolyChord} and can be found in any chain based sampler. 

The key piece lacking from the simulated weights method is that it uses the single value of $f(\theta)$ available for each iso-likelihood contour as a proxy for the average $f(\theta)$ over the contour, since there is no extra information in the live points to do otherwise. However, phantom points, drawn from the likelihood-constrained prior and each with a corresponding likelihood evaluation, have the potential to provide this additional information about the variation of $f(\theta)$ along contours.  Below, we present two novel uncertainty estimation methods which incorporate this. 

\subsection{Phantom points-informed uncertainty estimation}
\subsubsection{Method 1: Likelihood binning}\label{sec:likelihoodbinning}

In this approach, we bin the phantom points from the run by their likelihood values, such that each phantom point is assigned to the dead point to which it is closest in log-likelihood. Thus, each dead point is now associated with a set of points which sit very close to the contour defined by it; we make the assumption that, though the phantom points do not lie exactly on the dead point's iso-likelihood contour, they are still representative of the distribution of the $f(\theta)$ values along the contour. Then, for each dead point in our sum in equation~\ref{eq:pe_sum}, we resample a new $f(\theta)$ value from the associated bin, which includes the original dead point itself, as well as sampling a new $\Delta X$. The uncertainty estimate is obtained by repeating this process many times and taking the standard deviation of the resulting distribution of estimates. 

In reality, the phantom points are slightly correlated, which can lead to biased results due to the points not independently populating the prior. However, the large majority of the correlation can be removed by only using every, for example, $5^{\textrm{th}}$ point in the chains, as the correlation in the chirp mass becomes negligible after a few steps. This is discussed further in Section~\ref{binresults}.

\subsubsection{Method 2: Reconstructed runs}\label{sec:reconstructedruns}

In the second method, we first note that there is nothing wrong with any of the phantom points in terms of their suitability for use in a nested sampling run, except that we have already chosen to use another point (the associated dead point) which may be too correlated with the phantom point to use both. Hence, we may take the $1^{\textrm{st}}$ phantom point in every chain in the run and combine these carefully to form an equally valid nested sampling run to the original. Thus, from every run we are able to reconstruct multiple equally valid `phantom runs'. It is true that some of these runs will be correlated to each other, but, as discussed above, this correlation can be largely removed by only using a subset of the phantom points. 

These extra phantom runs are akin to performing multiple nested sampling runs on the same dataset, but, crucially for gravitational wave inference, at no extra computational cost. The parameter estimates from each of these reconstructed runs, as well as the original run, can then be combined and the corresponding new error bar may be computed from this. 

\subsubsection{Verifying accuracy of error bar}\label{verifymethod}

For certain parameters, the chain length of the sampler may not be long enough to accumulate enough uncorrelated phantom points that populate the prior in the correct way. In this case, the new error bars from the above two methods, though certainly wider than those from the usual method, may still not be wide enough to capture the full variation of the parameter along the contours. For gravitational waves, parameters like the mass ratio and the luminosity distance may suffer from this. If one is able to perform many nested sampling runs on the dataset, the error bars can be checked rigorously, as in Section~\ref{demo}. However, this is inefficient and in many cases simply not viable. We need to be able to verify from a single run whether the new error bars are now fully correct for a given parameter. We propose a method to check for convergent results of the parameter estimates, indicating whether the error bars have been quantified correctly.

For the default chain length in \textsc{PolyChord} the phantom points are very well distributed in their chirp mass values in order to be able to obtain an accurate error bar for this parameter using them. But they are more correlated in their luminosity distance values, meaning that the error bars obtained for this parameter will still be an underestimate of the true uncertainty. We need to employ a longer chain length in the run in order to obtain sufficient usable phantom points for the two methods described above. We can check whether there are enough uncorrelated phantom points for a given parameter by dividing the points into two halves, according to their position within the chains. We can then take each half separately and use either the likelihood binning method or the reconstructed runs method to obtain a set of parameter estimates, checking whether the two sets of estimates are in agreement using the Kolmogorov-Smirnov 2-sample test. We demonstrate this method on the chirp mass and the luminosity distance in Section~\ref{verify}.

\section{Results and Discussion}\label{results}

\subsection{Likelihood binning}\label{binresults}

\begin{figure}
    \centering
    \includegraphics{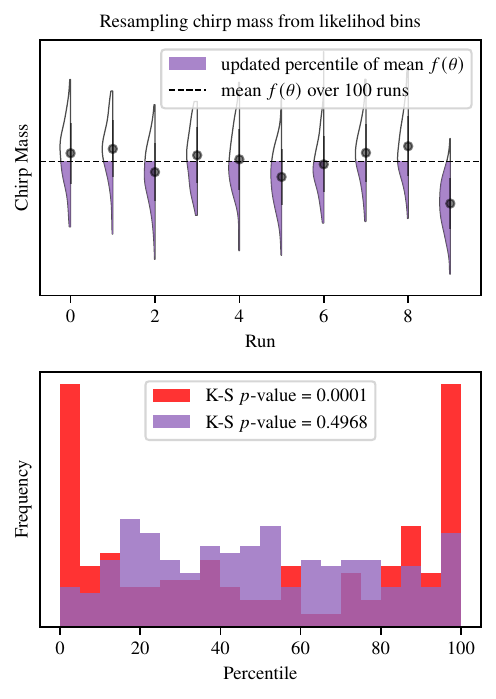}
    \caption{For each of the first 10 nested sampling runs performed on our simulated dataset, the new distribution of chirp mass estimates are plotted, obtained from resampling both shrinkage ratios and chirp mass values from likelihood bins around each contour (top). Resampling both the weights and the chirp mass gives wider error bars, as expected, and even by eye these seem more consistent with the spread of estimates across runs. Testing the likelihood binning method more rigorously, we see that the percentiles of the overall mean chirp mass are now consistent with being uniformly distributed (bottom, purple). This indicates that the true variation of the chirp mass along the contours is now being correctly accounted for, and thus the stochasticity of nested sampling parameter estimation is properly captured.}
    \label{fig:likelihood_binning}
\end{figure}

To demonstrate the efficacy of the likelihood binning method, we take the same 100 nested sampling runs used in Section~\ref{demo}, and apply the method as laid out in Section~\ref{sec:likelihoodbinning} to estimate the mean chirp mass of the simulated BBH event and its associated uncertainty. The results are shown in Figure~\ref{fig:likelihood_binning}. As in Figure~\ref{fig:combined_violins}, we include a histogram of the percentiles of the `true chirp mass', estimated from the mean chirp mass of the 100 runs. As demonstrated by the K-S test $p$-value, the distribution of percentiles is now much more consistent with a uniform distribution. This shows that the error bars now accurately capture the variation in the chirp mass over the contours, and reflect the run-to-run differences from nested sampling due to this additional stochasticity. 

In Section~\ref{sec:likelihoodbinning}, we suggested using only every $5^{\textrm{th}}$ phantom point in order to mitigate the slight correlation between successive phantom points in a chain. In general, how many phantom points should be skipped will depend on the exact sampler and settings, but there will likely be several sensible choices that give comparable results. The metrics provided in this paper, particularly in Sections~\ref{verifymethod} and~\ref{verify}, provide a way to decide on a suitable choice. For the \textsc{PolyChord} sampler and the settings we chose for these runs, we found every $5^\textrm{th}$ point to be a sensible choice.

\subsection{Reconstructed runs}\label{runresults}

\begin{figure}
    \centering
    \includegraphics{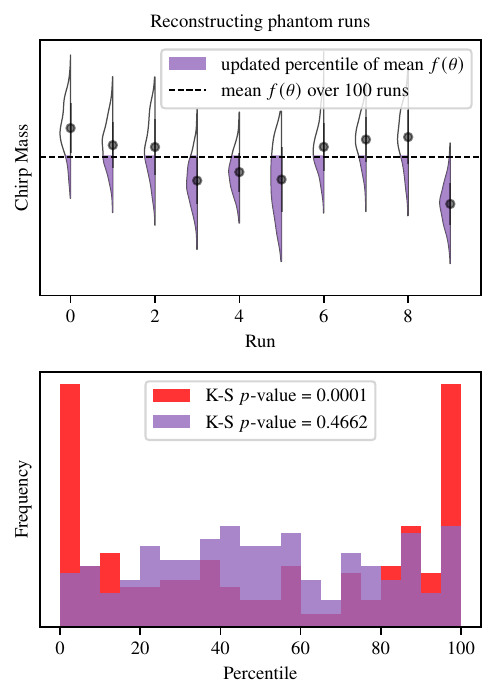}
    \caption{As with the likelihood binning method, computing parameter estimates from the reconstructed phantom runs gives correctly distributed percentiles. Again, we capture the stochasticity of nested sampling parameter estimation without additional computational cost.}
    \label{fig:reconstructed_runs}
\end{figure}

As above, we use the same set of 100 nested sampling runs, this time applying the method described in Section~\ref{sec:reconstructedruns}. The resulting percentiles plot is shown in Figure~\ref{fig:reconstructed_runs}, where it can be seen that, as with the first method, the new error bars are now much more consistent with the spread of estimates across multiple runs. The efficacy of these methods were also tested on other parameters besides the chirp mass, including parameters which were not explicitly sampled over. These results are included in Appendix~\ref{appendixA}. 

\subsection{Accuracy of new error bars}\label{verify}

\begin{figure}
    \centering
    \includegraphics{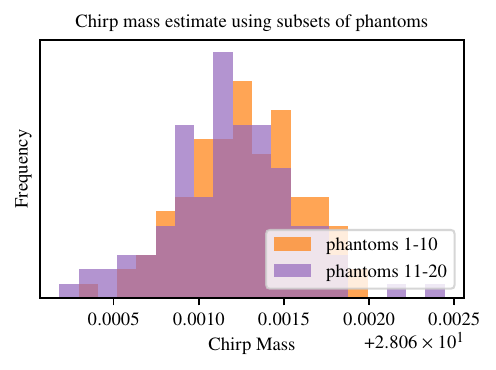}
    \caption{The chirp mass was estimated from a single run using the likelihood binning method described in Section~\ref{sec:likelihoodbinning}. Two sets of estimates were produced, one using only the first ten phantom points in all the chains and the other using only the last ten. The resulting distributions are in agreement with each other, with the K-S 2-sample test statistic being $0.13$, with a $p$-value of $0.37$. This indicates that the phantom points were well distributed in the chirp mass and give an accurate estimate of the true error bar. }
    \label{fig:chirpcheck}
\end{figure}

\begin{figure}
    \centering
    \includegraphics{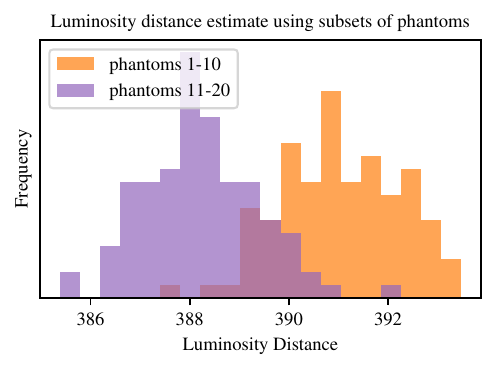}
    \caption{The same analysis was repeated for the luminosity distance, a parameter on which our methods do not perform well with the default \textsc{PolyChord} chain length. This time the K-S 2-sample test statistic was $0.77$, with a $p$-value of $1.92\times10^{-29}$. It is likely that the phantom points are still a little too correlated to their associated dead point in luminosity distance, and a longer chain length is needed as more of the points in the chain must be discarded.}
    \label{fig:lumdistcheck}
\end{figure}

We illustrate below the method proposed in Section~\ref{verify} to check whether the error bars from the likelihood binning method and the reconstructed runs method are indeed correct, without the need to perform multiple runs and plot a histogram of percentiles as above. From this section onwards, for conciseness all plots use the likelihood binning method only, but both methods perform similarly well on all considered examples.

The two uncertainty estimation methods demonstrated above perform very well on parameters such as the chirp mass, but underestimate the uncertainty on parameters such as the luminosity distance. We apply the approach proposed in Section~\ref{verify} to verify the error bars on both these parameters from a single run and obtain Figures~\ref{fig:chirpcheck} and~\ref{fig:lumdistcheck}. For the chirp mass, it can be seen that the estimates obtained from using only the first ten phantom points in the chain agree well with the estimates from the last ten phantom points in the chain. The same cannot be said for the luminosity distance, indicating that we may need a chain length longer than 20 in order to obtain enough uncorrelated phantom points to estimate the uncertainty on this parameter. It is important to note here that this does not mean the chain length of the run wasn't long enough to produce uncorrelated posterior samples in the luminosity distance, only that it was not long enough to produce sufficient suitable phantom points. 

\begin{figure}
    \centering
    \includegraphics{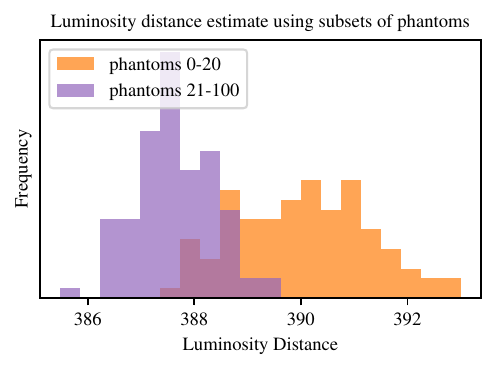}
    \caption{For a run with a chain length of $100$, the luminosity distance estimates from the first $20$ points in the chain are compared to those from the rest of the phantom points. The K-S 2-sample test statistic for the resulting set of estimates is $0.81$, with an accompanying $p$-value of $3.94\times10^{-33}$. This indicates that the first 20 phantom points are not well distributed enough to make an accurate estimate of the luminosity distance error bars and more points are needed. }
    \label{fig:lumdist_check1}
\end{figure}

\begin{figure}
    \centering
    \includegraphics{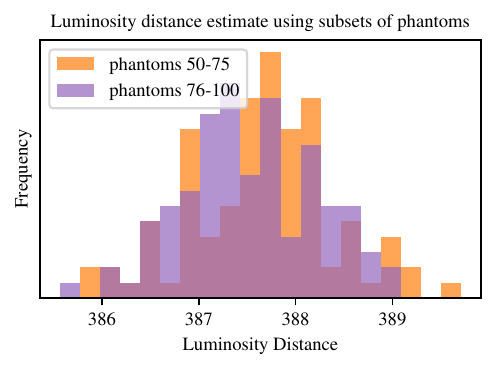}
    \caption{The same analysis is repeated, but comparing estimates obtained from the phantom points in the chains between positions 50 and 75 and the estimates from the just the last 25 points in each chain. These two sets of estimates now show good agreement, with a test statistic of $0.17$ and a $p$-value of $0.11$. Using the last 50 phantom points in the chains for either method described in this paper would therefore give an accurate estimate of the true luminosity distance error bar. }
    \label{fig:lumdist_check2}
\end{figure}

To examine this further, we performed a run with a chain length of $100$, to see how many phantom points must be discarded before we have enough uncorrelated points to provide convergent estimates of the luminosity distance. First, the estimates obtained from the first 20 phantom points in the chain are compared to those from the last 80 points, shown in Figure~\ref{fig:lumdist_check1}. The two sets of estimates are very different and this is a clear indication that the original chain length of $20$ was insufficient to accurately estimate the luminosity distance error bars. Next, the estimates obtained from phantoms points at positions in the chains between 50 and 75 were compared to estimates from the last 25 points in the chains. The two sets of estimates in Figure~\ref{fig:lumdist_check2} are now in agreement and show that the last 50 phantom points in the chain may be used in the likelihood binning or reconstructed runs method to provide an accurate estimate of the error bars. 

\subsection{Comparison to Higson method}\label{nestcheck}

\begin{figure}
    \centering
    \includegraphics{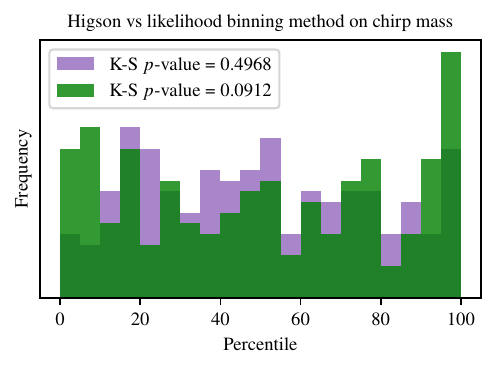}
    \caption{The Higson method (green) and the methods presented in this paper (purple) perform similarly well on the simulated BBH signals, producing percentile plots that are consistent with a uniform distribution. Both are a significant improvement on the usual `simulated weights' method, and are distinct in their approaches so may be used in conjunction with each other.}
    \label{fig:nestcheck}
\end{figure}

In order to compare the methods presented in this paper to those obtained using the bootstraps method presented in~\cite{Higson_2018}, we use the \texttt{nestcheck}~\citep{nestcheck} package to apply the Higson method on the same set of simulated BBH signals. In this method, a nested sampling run is split up into single live point runs, and then $n_\textrm{live}$ runs are sampled with replacement from these. This is repeated many times to build up a distribution of parameter estimates from which the error bar is estimated. 

Figure~\ref{fig:nestcheck} shows the resulting percentiles plot for the chirp mass; the Higson method performs similarly to the likelihood binning method and provides an independent way to correctly estimate the error bars. On parameters where the phantom methods do not work as well for the default settings, such as the luminosity distance, the Higson method also does not produce wide enough error bars. The advantage of using phantom points is that they can confirm from a single run whether or not the error bars are correct for a given parameter, as demonstrated in the previous section. 

The phantom methods are distinct from the Higson method, and so they can be used in combination. Using phantom points and single live point run bootstrapping together may also help to obtain the correct error bars for parameters like the luminosity distance without having to run a longer chain. 

\subsection{Coverage and credible intervals}\label{coverage}

The additional uncertainty in nested sampling parameter estimation described here affects not only our estimates of parameter means, but also the credible intervals. If a parameter inference procedure is accurate, we expect that the percentiles of the true parameters across many analysed datasets should be uniformly distributed~\citep{pplot1, pplot2}. The tool typically used to check this is a $p$-$p$ plot~\citep{ppcheck0,ppcheck1,ppcheck2,ppcheck3,ppcheck4,ppcheck5,ppcheck6,ppcheck7,ppcheck8, ppcheck9}, constructed from $N$ runs on injected signals where the true parameters are known. 

\begin{figure}
    \centering
    \includegraphics{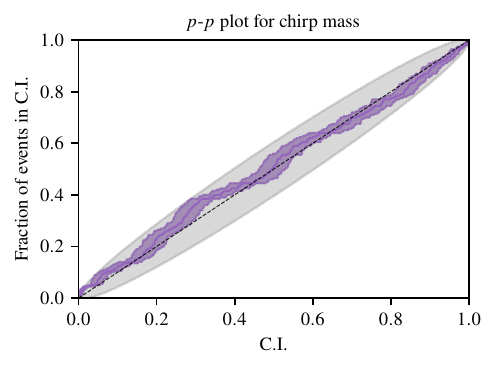}
    \caption{The $p$-$p$ plot obtained for the chirp mass using \texttt{bilby} and \textsc{PolyChord} is shown. As with the parameter mean estimates, we only sample over four parameters, simply setting the rest to the injected value. The plot shows the cumulative distribution function (CDF) of the percentiles of the true chirp mass for $200$ events, calculated in the usual way, with the associated 3-$\sigma$ confidence interval due to the statistical uncertainty from the finite number of events (gray). The purple region shows the 3-$\sigma$ confidence interval on the calculated CDF of the percentiles due to the stochastic nature of nested sampling. The binomial error still dominates, but the nested sampling parameter estimation uncertainty can be useful in assessing the source of deviations from the expected distribution.}
    \label{fig:pp-plot}
\end{figure}

\begin{figure}
    \centering
    \includegraphics{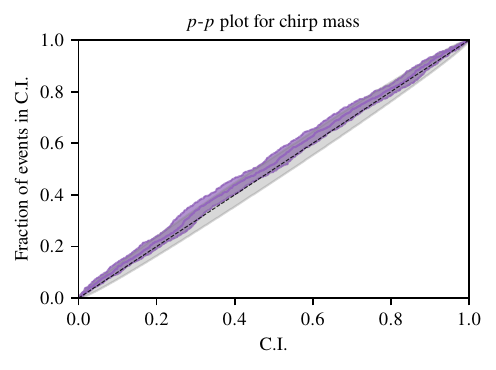}
    \caption{The CDF of the percentiles of the true chirp mass is shown for $1000$ events, along with the 3-$\sigma$ confidence interval due to a finite event sample size (gray). The 3-$\sigma$ confidence interval due to the nested sampling parameter estimation uncertainty is also shown (purple). These are now of a comparable size, and in order to make an accurate assessment of the coverage, it is necessary to account for both sources of uncertainties.}
    \label{fig:pp-plot-1000}
\end{figure}

A $p$-$p$ plot shows the fraction of events for which the true parameter values lie within a given credible interval as a function of the credible interval. In the ideal case where the parameter inference procedure leads to the correct coverage, this plot should be a diagonal line. There is a statistical uncertainty that arises from the finite number, $N$, of datasets analysed, often indicated by a gray region around the ideal diagonal line (as in Figures~\ref{fig:pp-plot} and~\ref{fig:pp-plot-1000}), and this is usually the dominant source of uncertainty in $p$-$p$ plot analyses. However, as $N$ increases, the nested sampling uncertainty becomes important. This uncertainty can be evaluated using either the likelihood binning or reconstructed runs method, and we can place a `nested sampling confidence interval' on our $p$-$p$ plot to account for the variation of the calculated percentiles from run to run on the same set of events. 

Figures~\ref{fig:pp-plot} and~\ref{fig:pp-plot-1000} were produced using the likelihood binning method. For each posterior sample from a given event, the individual parameter values were resampled from bins of neighbouring phantom points. These resampled parameter values were then used to calculate a set of credible intervals for that event, from which the uncertainty on the credible interval was estimated. In Figure~\ref{fig:pp-plot}, a total of $200$ injected signals were analysed, with the injection parameters drawn from the prior, and the gray region shows the 3-$\sigma$ ($99.7$\%) confidence interval for $N=200$ due to a finite event sample size. The purple region shows the corresponding 3-$\sigma$ confidence interval on the $p$-$p$ curve due to the nested sampling parameter estimation uncertainty. In scenarios where the $p$-$p$ plot lies outside of the gray region, using the nested sampling error bars may help to assess how much of the deviation is simply due to the stochastic nature of nested sampling. Figure~\ref{fig:pp-plot-1000} is produced from $1000$ injected signals, where now the statistical uncertainty for $N=1000$ and the nested sampling parameter estimation uncertainty are of similar sizes. Here, it is especially useful to consider the latter uncertainty in assessing to what extent the $p$-$p$ plot displays the expected coverage. Ideally, the gray confidence intervals should include both sources of uncertainties.

\section{Conclusions}\label{conc}

In nested sampling, the evidence error bar is dominated by the unknown prior volumes of the `shells' between iso-likelihood contours. This affects parameter estimation too, but here there is an additional source of uncertainty, from using the parameter value of a single sample, $f(\theta_i)$, as a proxy for the average parameter value over the entire iso-likelihood contour defined by that sample, $\tilde{f}(X_i)$~\citep{Chopin_Robert}. Though the prior volume uncertainty is well understood and accurately estimated from the `simulated weights method', the added stochasticity due to parameter variation over a given contour is typically ignored, despite the fact that this can be a significant source of uncertainty in the analysis of any dataset, including gravitational wave signals~\citep{Thrane_2019}. 

Here we have proposed two novel approaches to account for this additional uncertainty using the extra likelihood calls made at run-time by any chain based sampler. Though these `phantom points' are not suitable for use in evidence estimation, they are valuable in parameter inference for exploring the variation in a parameter value over individual contours. This is the first demonstration of the use of nested sampling phantom points for inference. We have also shown how to use phantom points to verify the accuracy of the error bars on a given parameter, by splitting the set of the phantom points in two and checking for convergence in the resulting estimates. This could be a useful approach at runtime in tuning the chain length of chain based samplers to speed up nested sampling whilst still ensuring independent samples. In other MCMC procedures, such as Hamiltonian Monte Carlo~\citep{stan, blackjax}, the emphasis is placed on tracking convergence on specific parameters of interest, and dynamic nested sampling~\citep{higson_dynamicns, dynesty} provides a way to apply this in nested sampling too.  

Furthermore, we have shown how this uncertainty impacts estimates of credible intervals and coverage plots, such as the $p$-$p$ plot commonly used in gravitational wave analyses. Though this source of uncertainty is often secondary to the statistical uncertainty due to the finite number of events sampled for a typical coverage plot, the two become comparable in size for larger numbers of analysed events. This highlights the importance of accounting for the additional nested sampling parameter estimation uncertainty when interpreting such plots.

The methods described in this paper differ from the \texttt{nestcheck} methods proposed in~\cite{Higson_2018} in that we incorporate information from the chain based phantom points and do not make use of bootstraps, but they produce comparable results and should be viewed as complementary. They can be combined and used together. This may be particularly helpful for parameters where the original chain length was not long enough to produce sufficiently many uncorrelated phantom points for use with the methods in this paper on their own, and may avoid the need to repeat the run with a longer chain. The advantage of incorporating phantom points is that they can be used to assess whether the true run-to-run variation has been captured from a single run, in contrast with bootstrapping. 

Though phantom points are specific to chain based samplers, other samplers may generate discarded but valid points. These points also would not be suitable for evidence estimation but may be of use for other purposes such as parameter inference. Both importance sampling methods~\citep{nessai, nessai2, nautilus} and methods using machine learning proxies to accelerate nested sampling~\citep{bambi, nautilus} generate samples that are not used, but may provide additional information about the parameter space. Likewise, \textsc{MultiNest}~\citep{multinest_paper} also produces discarded points. We hope that this work will renew interest within the community~\citep{nature_ns_review} in methods for quantifying nested sampling parameter estimation. 

\section*{Acknowledgements}

MP was supported by the Harding Distinguished Postgraduate Scholars Programme (HDPSP). WH was supported by a Royal Society University Research Fellowship. This work was performed using the Cambridge Service for Data Driven Discovery (CSD3), part of which is operated by the University of Cambridge Research Computing on behalf of the STFC DiRAC HPC Facility (www.dirac.ac.uk). The DiRAC component of CSD3 was funded by BEIS capital funding via STFC capital grants ST/P002307/1 and ST/R002452/1 and STFC operations grant ST/R00689X/1. DiRAC is part of the National e-Infrastructure.

\section*{Data Availability}

All the data used in this analysis, including the relevant nested sampling dataframes, can be obtained from~\cite{zenodo}. We also include a notebook with all the code to reproduce the plots in this paper.



\bibliographystyle{mnras}
\bibliography{nspe} 




\appendix

\section{Other parameters}\label{appendixA}

As well as the chirp mass and luminosity distance, the methods presented in this paper were used to estimate uncertainties on other parameters. Since both methods perform similarly well to each other on all the parameters, for conciseness we present the plots for the likelihood binning method only.

\begin{figure}
    \centering
\includegraphics{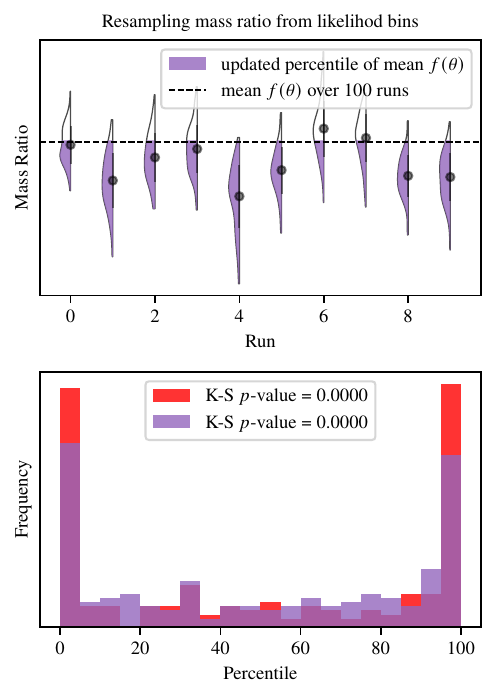}
    \caption{The mass ratio is an example of a parameter on which our methods (purple) did not significantly outperform standard methods (red) of uncertainty estimation for the default chain length settings we used. As with the luminosity distance, this can be rectified by setting the chain length to be longer.}
    \label{fig:massratio_violins}
\end{figure}

\begin{figure}
    \centering
    \includegraphics{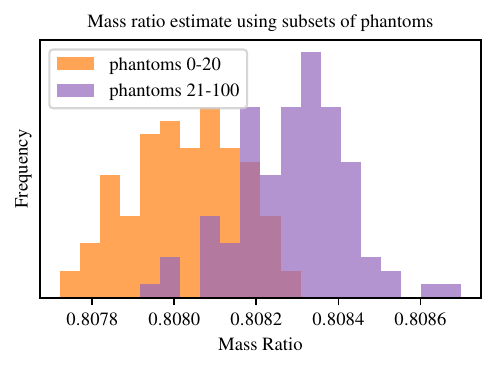}
    \caption{For a run with a chain length of $100$, the parameter estimates using the likelihood binning method from the first $20$ points in the chain do not match those from the last $80$. This indicates that the first $20$ points are perhaps too correlated in their mass ratio values to use in these methods.}
    \label{fig:massratio0to20}
\end{figure}

\begin{figure}
    \centering
    \includegraphics{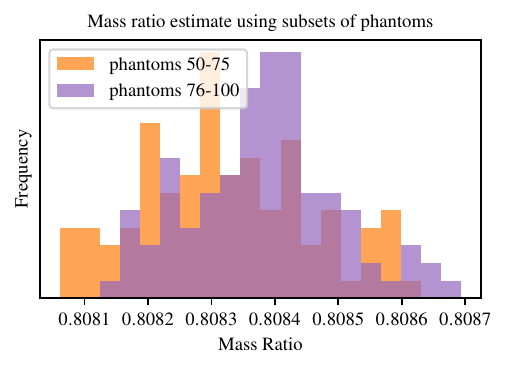}
    \caption{Running a longer chain corrects the issue. We can safely use the phantoms points from position $50$ onwards in the chain, and our methods will give the correct bars.}
    \label{fig:massratio50to75}
\end{figure}

As discussed above, on some parameters neither method works significantly better than the standard methods, and this is largely driven by the lack of sufficient uncorrelated phantom points in these parameters. The mass ratio (Figure~\ref{fig:massratio_violins}) is one such parameter. This can be rectified by running a longer chain during sampling. As discussed in Section~\ref{verify}, we can verify that the error bar on the mass ratio parameter computed from a single run with the default chain length is not correct by following the method described in Section~\ref{verifymethod}. Performing a run with a chain length of $100$ instead, Figures~\ref{fig:massratio0to20} and~\ref{fig:massratio50to75} corroborate that the original chain length of $20$ was too short to produce sufficiently many uncorrelated phantom points for use in these methods, and a much longer chain must be used.  

\begin{figure}
    \centering
    \includegraphics{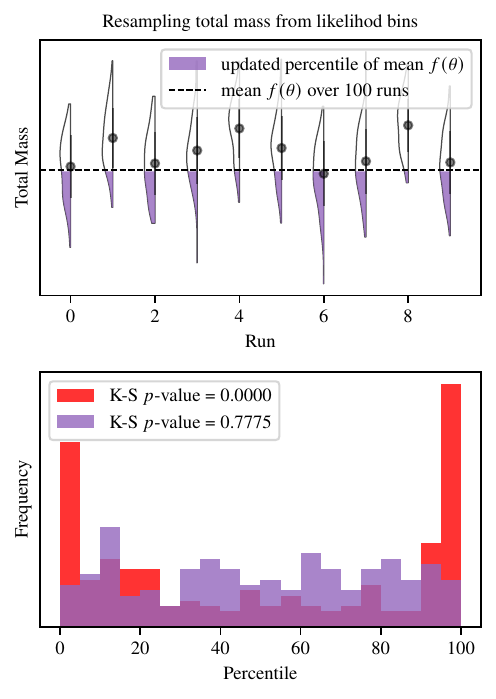}
    \caption{The total mass is a function of two sampled parameters, the chirp mass and the mass ratio. Our methods (purple) are able to successfully capture the true uncertainty on such a parameter, whereas the simulated weights method (red) is not.}
    \label{fig:totalmass_violins}
\end{figure}

However, on other parameters such as the total mass, both methods successfully calculate the correct error bars, as they did with the chirp mass (Figure~\ref{fig:totalmass_violins}). Crucially, the total mass is not one of the four sampled parameters in the run, and thus demonstrates the robustness of these methods to arbitrary functions of the sampling parameters. 

\section{Other samplers and settings}

The methods presented in this paper apply to any chain-based sampler, but the results may be dependent on the exact sampler settings used. Below, we use \textsc{PolyChord} with the default settings for slice sampling in \texttt{dynesty}~\citep{dynesty} to demonstrate that these settings need to be carefully decided using the metrics presented in this paper in order to obtain valid uncertainty estimates.  

\begin{figure*}
    \centering
    \includegraphics{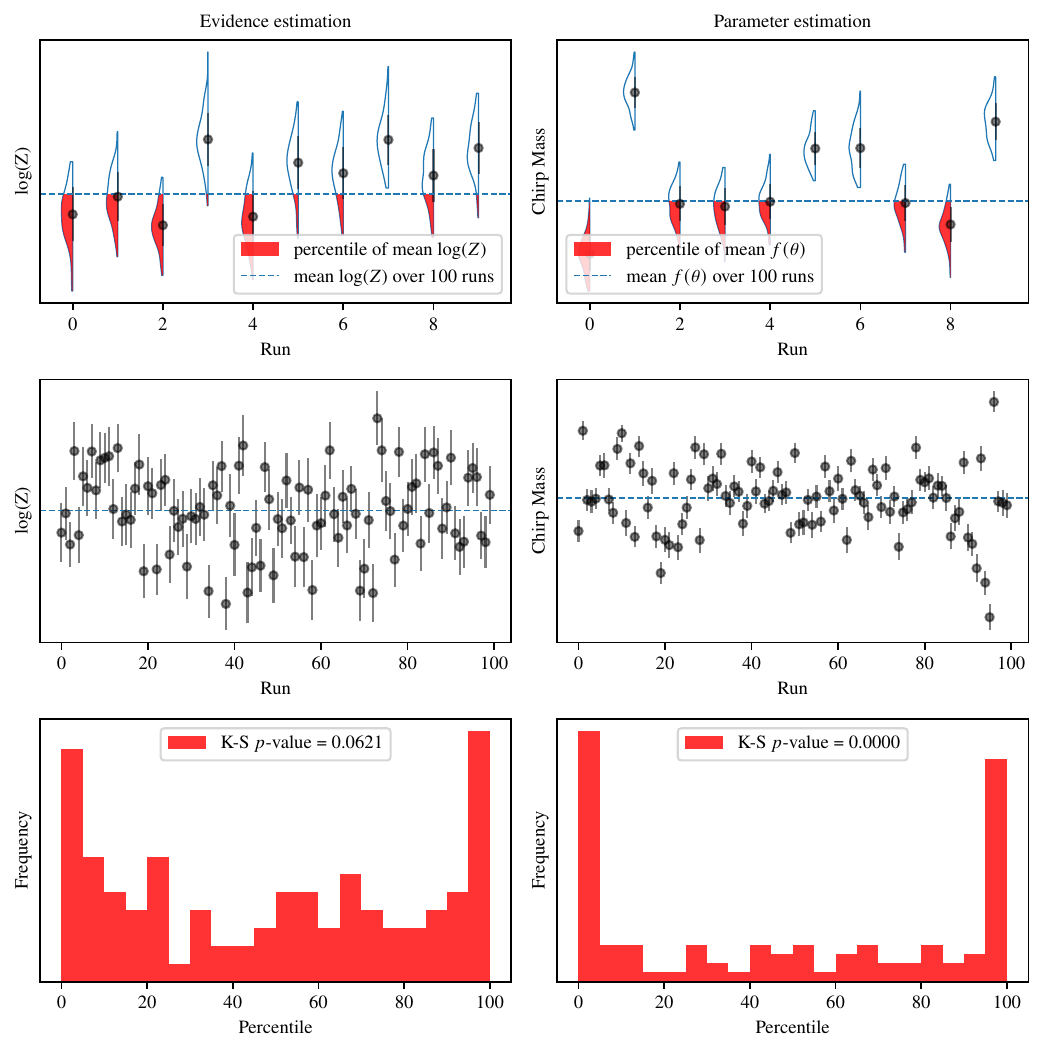}
    \caption{The equivalent of Figure~\ref{fig:combined_violins} for the new settings. The middle left panel shows the estimated error bars on the evidence using the `simulated weights' method and the bottom left shows a histogram of the percentiles in which the `true evidence', estimated over 100 runs, lies. The K-S test $p$-value for a uniform distribution is $0.0621$. On the right hand side panels, the equivalent figures for the chirp mass estimates are shown. These show qualitatively the same results as for the longer chain.}
    \label{fig:dynesty_evidence}
\end{figure*}

\begin{figure}
    \centering
    \includegraphics{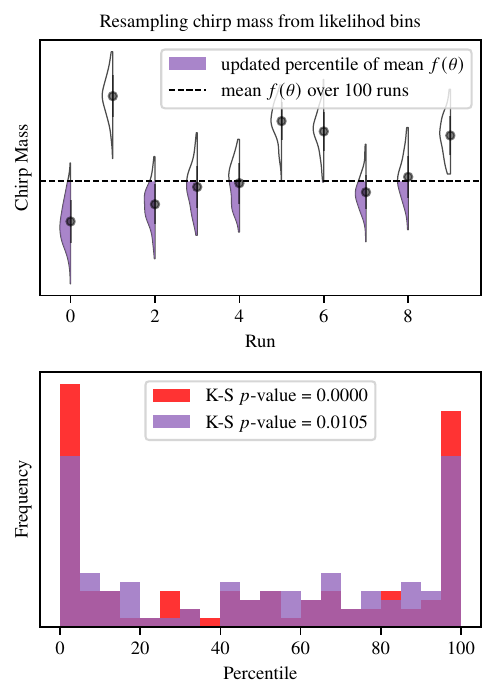}
    \caption{The likelihood binning method applied to these phantom points results in a distribution that is still not uniform, though the error bars are wider. This is due to the chain length not being long enough to produce sufficiently many uncorrelated phantom points.}
    \label{fig:dynesty_likelihoodbinning}
\end{figure}

\begin{figure}
    \centering
    \includegraphics{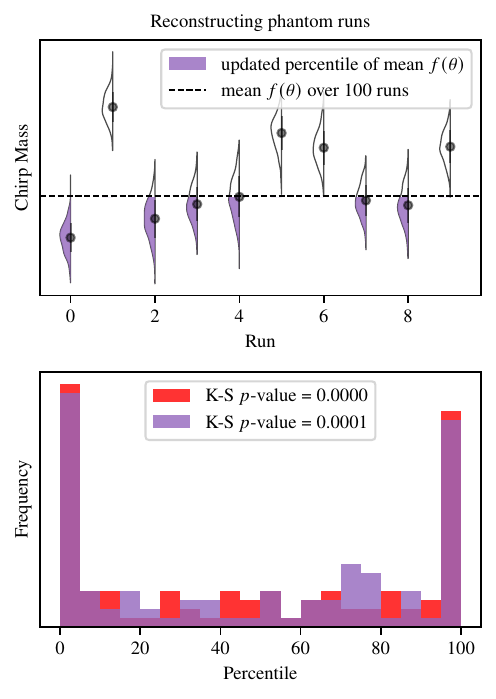}
    \caption{The phantom runs method also does not significantly improve the uncertainty estimate on the chirp mass, and here performs slightly worse than likelihood binning. This is again due to the chain length being too short.}
    \label{fig:dynesty_phantomruns}
\end{figure}

\begin{figure}
    \centering
    \includegraphics{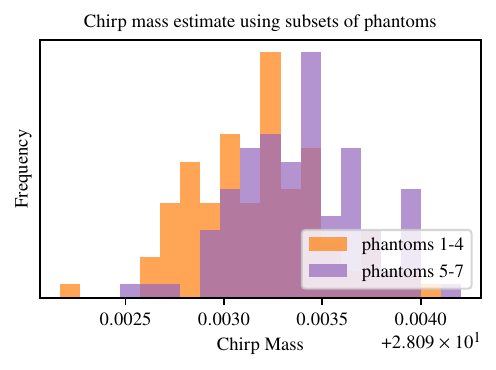}
    \caption{In order to test whether the chain length setting of $3 + \textrm{ndims}$, the default for slice sampling in \texttt{dynesty}, is long enough to apply these methods successfully, we can follow the methods presented in Section~\ref{verifymethod}. Splitting the phantom points into two sets, according to their position in the chains, we compare the estimates produced by each set independently. In this case, the K-S 2-sample test statistic between the resulting distributions was $0.31$, with a $p$-value of $2.2 \times 10^{-4}$.}
    \label{fig:dynesty_verify}
\end{figure}

For the slice sampler setting in \texttt{dynesty}, the default chain length is $3 + \textrm{ndims}$, in contrast to \textsc{PolyChord}'s $5 \times \textrm{ndims}$. We note here that \texttt{dynesty} does not default to slice sampling until the number of dimensions is above 20. However, the default sampler for 4 dimensions is not a chain-based one and thus our methods would not apply. Additionally, many users may choose correctly to change that default as rejection sampling can become inefficient in much lower dimensions too. \texttt{dynesty} also doesn't have a clustering algorithm as part of its nested sampler, and so to emulate this code, we run \textsc{PolyChord} with a chain length of $7$ for our $4$ parameter problem and turn off clustering in the settings. The corresponding results are presented in Figures~\ref{fig:dynesty_evidence}-~\ref{fig:dynesty_verify}. 

Neither the likelihood binning nor the phantom runs method works particularly well here, and this is because although the chain length may be long enough to produce unbiased estimates of the evidence, there aren't sufficiently many uncorrelated phantom points to use. We can verify this by following the method discussed in Section~\ref{verifymethod}, comparing the estimates obtained from the first 4 phantom points in every chain to those from the last 3. The resulting distributions have a K-S 2-sample test statistic of $0.31$, with a $p$-value of $2.2 \times 10^{-4}$, indicating that they may not be in agreement. This is a sign that a longer chain needs to be run.

The two methods in this paper should work for any chain-based sampler, but this example shows the importance of the chain length setting to the validity of the results. In order to get accurate parameter estimation uncertainties, it may be necessary to run a longer chain than is needed for unbiased evidence estimates. Again, the metrics presented in Section~\ref{verifymethod} will enable users to manually set the appropriate chain length. 


\bsp	
\label{lastpage}
\end{document}

%% file: schematic_revised_latex_big.pdf_tex
\begingroup%
  \makeatletter%
  \providecommand\color[2][]{%
    \errmessage{(Inkscape) Color is used for the text in Inkscape, but the package 'color.sty' is not loaded}%
    \renewcommand\color[2][]{}%
  }%
  \providecommand\transparent[1]{%
    \errmessage{(Inkscape) Transparency is used (non-zero) for the text in Inkscape, but the package 'transparent.sty' is not loaded}%
    \renewcommand\transparent[1]{}%
  }%
  \providecommand\rotatebox[2]{#2}%
  \newcommand*\fsize{\dimexpr\f@size pt\relax}%
  \newcommand*\lineheight[1]{\fontsize{\fsize}{#1\fsize}\selectfont}%
  \ifx\svgwidth\undefined%
    \setlength{\unitlength}{595.27559055bp}%
    \ifx\svgscale\undefined%
      \relax%
    \else%
      \setlength{\unitlength}{\unitlength * \real{\svgscale}}%
    \fi%
  \else%
    \setlength{\unitlength}{\svgwidth}%
  \fi%
  \global\let\svgwidth\undefined%
  \global\let\svgscale\undefined%
  \makeatother%
  \begin{picture}(1,1.41428571)%
    \lineheight{1}%
    \setlength\tabcolsep{0pt}%
    \put(0,0){\includegraphics[width=\unitlength,page=1]{schematic_revised_latex_big.pdf}}%
    \put(0,0){\includegraphics[width=\unitlength,page=2]{schematic_revised_latex_big.pdf}}%
    \put(0.00195032,0.91203618){\color[rgb]{0,0,0}\makebox(0,0)[lt]{\lineheight{1.25}\smash{\begin{tabular}[t]{l}$\theta_1$\end{tabular}}}}%
    \put(0.47572936,0.50490559){\color[rgb]{0,0,0}\makebox(0,0)[lt]{\lineheight{1.25}\smash{\begin{tabular}[t]{l}$\theta_2$\end{tabular}}}}%
    \put(0.16551125,1.18648457){\color[rgb]{0,0,0}\makebox(0,0)[lt]{\lineheight{1.25}\smash{\begin{tabular}[t]{l}$\mathcal{L}_1$\end{tabular}}}}%
    \put(0.41587764,0.7634082){\color[rgb]{0,0,0}\makebox(0,0)[lt]{\lineheight{1.25}\smash{\begin{tabular}[t]{l}$\mathcal{L}_2$\end{tabular}}}}%
    \put(0.43714546,1.03231323){\color[rgb]{0,0,0}\makebox(0,0)[lt]{\lineheight{1.25}\smash{\begin{tabular}[t]{l}$\mathcal{L}_3$\end{tabular}}}}%
  \end{picture}%
\endgroup%